\documentclass[aps,prb,showpacs,notitlepage,reprint]{revtex4-1}
\usepackage{graphicx}
\usepackage{dcolumn}
\usepackage{amsmath}
\usepackage{bm}
\usepackage{todonotes}
\usepackage[utf8]{inputenc}
\usepackage{amsmath}
\usepackage{natbib}
\usepackage{graphicx}
\usepackage{caption}
\usepackage{subcaption}
\usepackage{lipsum}
\usepackage{float}

\newcommand{\bn}[1]{\mbox{\boldmath$#1$}}

\newcommand{\beq}{\begin{equation}}
\newcommand{\eeq}{\end {equation}}
\newcommand{\bea}{\begin{eqnarray}}
\newcommand{\eea}{\end{eqnarray}}

\begin{document}
\title{Direct Observation of the Faraday Rotation Using Radially-Polarized Twisted Light}
\author{F. Tambag$^{1}$, {\rm {K. Koksal}}$^{1}$, F. Yildiz$^{3}$, M. Babiker$^{4}$ } 
\affiliation{$^1$Physics Department, Bitlis Eren University, Bitlis, Turkey}
\affiliation{$^{3}$Department of Physics, Gebze Technical University, Kocaeli, Turkey}
\affiliation{$^{4}$Department of Physics, University of York, YO10 5DD, UK} 
\date{\today}

\begin{abstract}

A novel experimental technique  for the realisation of the optical Faraday effect using Laguerre-Gaussian (LG) light is described. The experiment employs a zero-order vortex half-wave retarder to generate a radially or azimuthally-polarised LG doughnut beam.  The light emerging from the retarder then passes through a linear polariser, which gives rise to two intensity lobes, with the orientation of the intensity gap between the two lobes pointing  parallel  (perpendicular) to the polarization direction of the radially (azimuthally) polarised beam. To complete the Faraday set up, the light traverses a material subject to a magnetic field, before passing through a final linear polariser, which
results in a visible rotation of the lobes pattern. This technique exhibits the Faraday effect readily visually, without further elaborate steps to detect changes in the light intensity.  The degree of rotation of the plane of polarisation is determined directly by the visibly clear change in the orientation of the intensity gap between the lobes.
\end{abstract}


\maketitle
 
There is currently a revival of interest in the optical Faraday effect fuelled by new applications, as for example, in the search for suitable magneto-active materials for Faraday isolators operating at different wavelengths \cite{vojna2020} and for small magnetic field measurements \cite{vle2019}.  

The Faraday effect is a well  known long established phenomenon exhibited by polarised light beams passing through slabs of magnetised materials.  Besides interest in it as a fundamental physical phenomenon, it had led to earlier useful applications in physics \cite{schatz1969faraday,bennett1965faraday,serber1932theory,boswarva1962faraday}. 
It manifests itself as a rotation of the sense of wave polarization caused by an axial magnetic field in the presence of a material and its magnitude depends on the Verdet constant \cite{vojna2019verdet,vojna2020faraday}, the length of material and the strength of the applied magnetic field. The traditional technique that is normally used to exhibit the Faraday rotation is to position a linear polariser after the beam has passed through the material and observe the change in the intensity of the light using photo-detectors \cite{jenkins1957fundamentals}. 
\begin{figure*}[ht]
\begin{subfigure}{1\columnwidth}
  \centering
  \includegraphics[width=.95\linewidth]{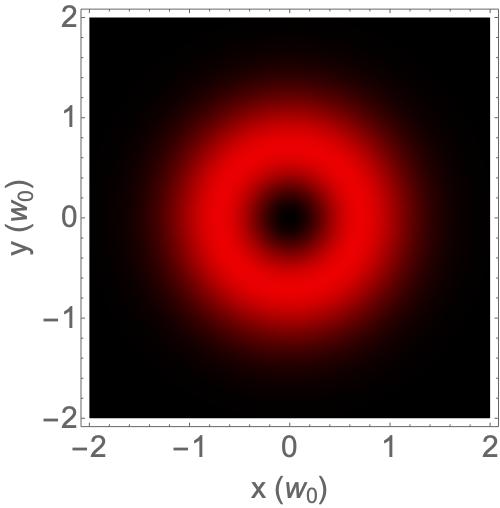}  
  \caption{Intensity profile of radially polarized LG beam without polarizer, ${\cal I}_{rad}$.}
  \label{fig:sub-firsta}
\end{subfigure}
\begin{subfigure}{1\columnwidth}
  \centering
  \includegraphics[width=.95\linewidth]{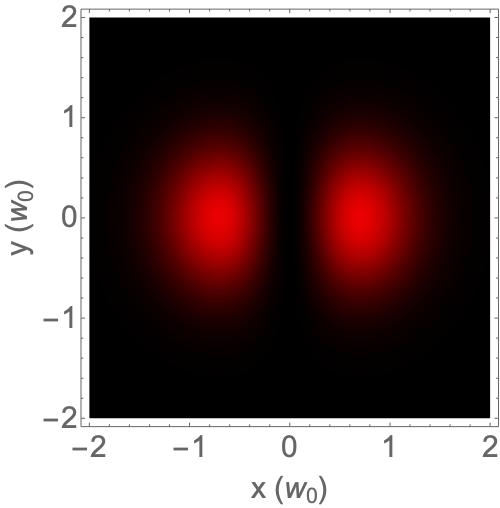}  
  \caption{Intensity profile of radially polarized LG beam after linear polarizer, ${\cal I}_{rad'}$.}
  \label{fig:sub-seconda}
\end{subfigure}
\caption{The doughnut shape intensity profile of the radially-polarized (or azimuthally-polarized) LG beam is converted into two lobes (as seen in Eq. \ref{vecx}) by using a linear polarizer.} 
\label{fig:profile}
\end{figure*} 

In this article, we put forward and describe a different technique experimentally and theoretically for the realisation of the Faraday effect which involves the use of radially or azimuthally-polarized Laguerre-Gaussian (LG) beams \cite{allen1999,andrews2012}. Besides its novelty due to the use of twisted light and direct observation of rotation angles, this technique has the advantage of eliminating the  use of  photo-detectors and the associated software which calculates the change in the light intensity in order to exhibit the Faraday rotation  \cite{loeffler1983faraday,pedrotti1990faraday,berman2010optical}. 
Radially and azimuthally polarized LG beams can be produced \cite{olvera2020cylindrically} by using a zero order vortex half-wave retarder \footnote{Thorlabs zero-order vortex half-wave retarder, model WPV$10$}. 

We focus on the case of a radially-polarised paraxial LG beam of frequency $\omega$ and axial wavevector $k$ travelling along the z-direction in a medium of refractive index $n$ and we clarify how the procedure can be applied to the case of an azimuthally-polarised LG mode. In cylindrical coordinates ${\bf r}=(\rho,\phi,z)$ a radially-polarised LG mode the electric field is written in the form 
\beq 
{\bf E}_{\rho}({\bf r},t)={\cal U}_0{\cal F}_{\ell p}(\rho)e^{i(nkz-\omega t)}e^{i\ell\phi}  {\bn {\hat \rho}}
\label{erad}
\eeq
where ${\cal U}_0$ is a constant;  ${\cal F}_{\ell p}(\rho)$ is the amplitude function of the paraxial Laguerre-Gaussian mode.
Explicitly for an LG$_{\ell p}$ mode the amplitude function ${\cal F}_{\ell p}$ is as follows 
\beq
{\cal F}_{\ell p}(\rho)= e^{-\frac{\rho^2}{w_0^2}}  \left(\frac{\sqrt{2} \rho}{w_0}\right)^{|\ell| }L^{|\ell|}_p\left(\frac{2\rho^2}{w_0^2}\right)
\label{LG}
\eeq
Here $L_p^{|\ell|}$ is the associated Laguerre polynomial of indices $|\ell|$ and $p$. The intensity of the radially-polarised beam is then given by 
\beq
{\cal I}_{\rho}=\frac{1}{2}\epsilon_0 c{\cal U}_0^2|{\cal F}_{\ell p}(\rho)|^2
\eeq
For p=0, this is a doughnut shape, as shown in Fig.\ref{fig:sub-firsta}.

The radially-polarized unit vector can be expressed in terms of circular polarisations ${\sigma_{\pm}}$ as follows  
\bea
{\bf E}_{\sigma_-}=e^{i(\ell+1)\phi}({\bn {\hat x}}-i{\bn {\hat y}}){\cal U}_0 {\cal F}_{\ell p}(\rho)e^{i(nkz-\omega t)}\\
{\bf  E}_{\sigma_+}=e^{i(\ell-1)\phi}({\bn {\hat x}}+
i{\bn {\hat y}}){\cal U}_0 {\cal F}_{\ell p}(\rho) e^{i(nkz-\omega t)}
\eea
so that
\beq
{\bf E}_{\rho}={\bf E_{\sigma_-}}+{\bf E_{\sigma_+}}
\eeq
Now we assume that the radially-polarised beam just created is made to pass through a linear polariser. After passing through the linear polarizer the  electric field vector of the two components ${\sigma_{\pm}}$ of the originally radially-polarized mode both become polarised along ${\bn {\hat x}}$ and we have
\beq
{\bf E}_{\rho'}=({\bf  E_{\sigma_-}}\longrightarrow{\bf  E_{x_1}})+({\bf  E_{\sigma_+}}\longrightarrow{\bf E_{x_2}})
\label{int}
\eeq
where ${\bf  E}_{x_1}$ and ${\bf  E}_{x_2}$ are given by
\beq
{\bf E}_{x_1}=e^{i(\ell+1)\phi}{\cal U}_0 {\cal F}_{\ell p}(\rho)e^{i(nkz-\omega t)}{\bn {\hat x}}
\eeq
\beq
{\bf E}_{x_2}=e^{i(\ell-1)\phi}{\cal U}_0 {\cal F}_{\ell p}(\rho) e^{i(nkz-\omega t)}{\bn {\hat x}}
\eeq
The total $x$-polarised electric field is the sum
\bea
{\bf E}_{\rho'}&=&\left(e^{i(\ell-1)\phi}+e^{i(\ell+1)\phi}\right){\cal U}_0 {\cal F}_{\ell p}(\rho) e^{i(nkz-\omega t)}{\bn {\hat x}}\nonumber\\
&=&(2 \; \cos\phi) \;{\cal U}_0 {\cal F}_{\ell p}(\rho) e^{i(nkz-\omega t)}e^{i\ell\phi}{\bn {\hat x}}
\label{int}
\eea
and we should note the appearance of the $\cos\phi$ factor in the electric field which we now see is the reason for the existence of two lobes in the intensity ${\cal I}'(\rho,\phi)$ of the light emerging from the linear poalriser which is given by
\bea
{\cal I}'(\rho,\phi)=2\epsilon_0 c{\cal U}_0^2|{\cal F}_{\ell p}(\rho)|^2 {\cos^2\phi}
\label{vecx}
\eea
It is easy to show that following the above procedure for the case of an  azimuthally-polarized LG mode will result in the ${\sin^2\phi}$ factor in the intensity being replaced by a ${\cos^2\phi}$ 
and the intensity pattern will then have a horizontal gap perpendicular to that for the radially-polarized case.

Figure \ref{fig:profile} displays the intensity distribution in the transverse plane of the two types of LG beams, namely, the radially-polarised beam and the linearly-polarised beam.  The well known doughnut shape profile of the radially-polarised LG beam is shown in Fig. \ref{fig:sub-firsta}. Fig.  \ref{fig:sub-seconda} displays the intensity profile of the beam 
which arises after the the original beam has passed through a linear polarizer, assuming that polarization direction of polarizer is along $x$. 
 
The theoretical basis of the optical Faraday experiment using twisted light follows similar lines to those put forward by  Berman \cite{berman2010optical} in the case of non-vortex light
with the only difference being that the cylindrical coordinate system is appropriate here for the description of the vortex beam and its effects.  

Once the radially-polarised LG beams enters the material where the magnetic field is applied, its electric field acquires an additional field component along the ${\bn {\hat \phi}}-$ direction, which is responsible for the Faraday rotation.  This electric field vector within the material is now as follows 
\beq
{\bf E}=E_{\rho} {\bn {\hat \rho}}+E_{\phi} {\bn {\hat \phi}}
\label{vecF2}
\eeq
where $E_{\rho}$ and $E_{\phi}$  are now given by
\bea
E_{\rho}&=&{\cal U}_0{\cal F}_{\ell p}(\rho)e^{i(nkz-\omega t)}e^{i\ell\phi}\label{vecF1}\\
E_{\phi}&=&\Theta(z)E_\rho
\label{efields}
\eea
The fields inside the magnetised material differ due to the additional component along ${\bn {\hat\phi}}$ which is proportional to  $\Theta(z)$.  This z-dependent angle stands for the Faraday angle of rotation  experienced by the beam within a length element $(dz)$ at axial position $z$ in the material.  

The total electric field vector when the beam emerges from the magnetised material is then given by 
\beq
{\bf E}={\cal U}_0{\cal F}_{\ell p}(\rho)e^{i(kz-\omega t)}e^{i\ell\phi}\left({\bn {\hat \rho}}+\Theta(z){\bn {\hat \phi}}\right)
\label{vecFrad}
\eeq
After exiting from the material the light beam enters a linear polariser (called the analyser) with the polarisation oriented along the x-direction, so it only selects the x-component of the electric field.  Expressing the unit vectors ${\bn {\hat\rho}}$ and ${\bn {\hat\phi}}$ in terms of ${\bn {\hat x}}$ and ${\bn {\hat y}}$ using the relations
\beq
{\bn {\hat\rho}}={\bn {\hat x}}\cos\phi-{\bn {\hat y}}\sin\phi;\;\;\;\;{\bn {\hat\phi}}={\bn {\hat x}}\sin\phi+{\bn {\hat y}}\cos\phi\label{rhophi}
\eeq
The form of electric field vector after leaving the final linear polariser (analyser) is obtainable from Eq.(\ref{vecFrad}) by substituting from Eqs.(\ref{rhophi}) and selecting the x-component, to be referred to as ${\bf E}_f$, which is  
\beq
{\bf E}_f={\cal U}_0{\cal F}_{\ell p}(\rho)e^{i(kz-\omega t)}e^{i\ell\phi}\left(\cos\phi+\Theta_F\sin\phi\right){\bn {\hat x}}
\label{vecFradxx}
\eeq
where $\Theta_F$ is the Faraday rotation angle; it is the cumulative angle when the beam traverses the full length $L$ of the magnetised material.
This angle is proportional to the product $B_0 L$ with the characteristic constant factor $V$, depending on the type of material
\beq
\Theta_F=VB_0L
\label{thetaeff}
\eeq
 
The light intensity corresponding to Eq.(\ref{vecFradxx}) is
\beq
{\cal I}=\frac{1}{2}\epsilon_0 c{\cal U}_0^2|{\cal F}_{\ell p}(\rho)|^2\left(\cos\phi+\Theta_F\sin\phi\right)^2 
\label{vecxrad1}
\eeq

\begin{figure*}[ht]
  \centering
  \includegraphics[width=.95\linewidth]{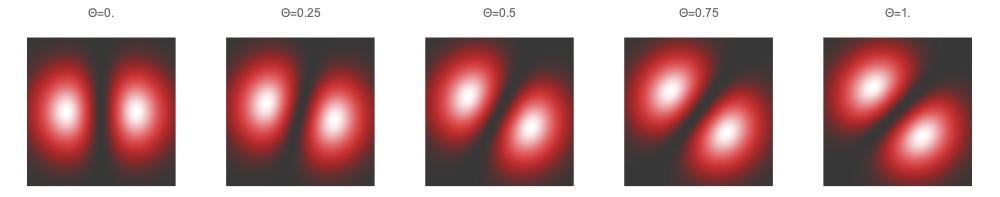}
\caption{The intensity profile corresponding to Eq. \ref{vecxrad1} showing the changes when varying the Verdet constant $\Theta_F$.  For $\Theta_F=0$, the intensity profile reduces to Eq. (\ref{vecx}). As in Eq. \ref{thetaeff}, the angle $\Theta_F$ depends on the applied magnetic field and the thickness $L$ of the material through which the beam passes}. 
\label{fig:theoretical1}
\end{figure*}
\begin{figure*}
  \centering
  \includegraphics[width=.8\linewidth]{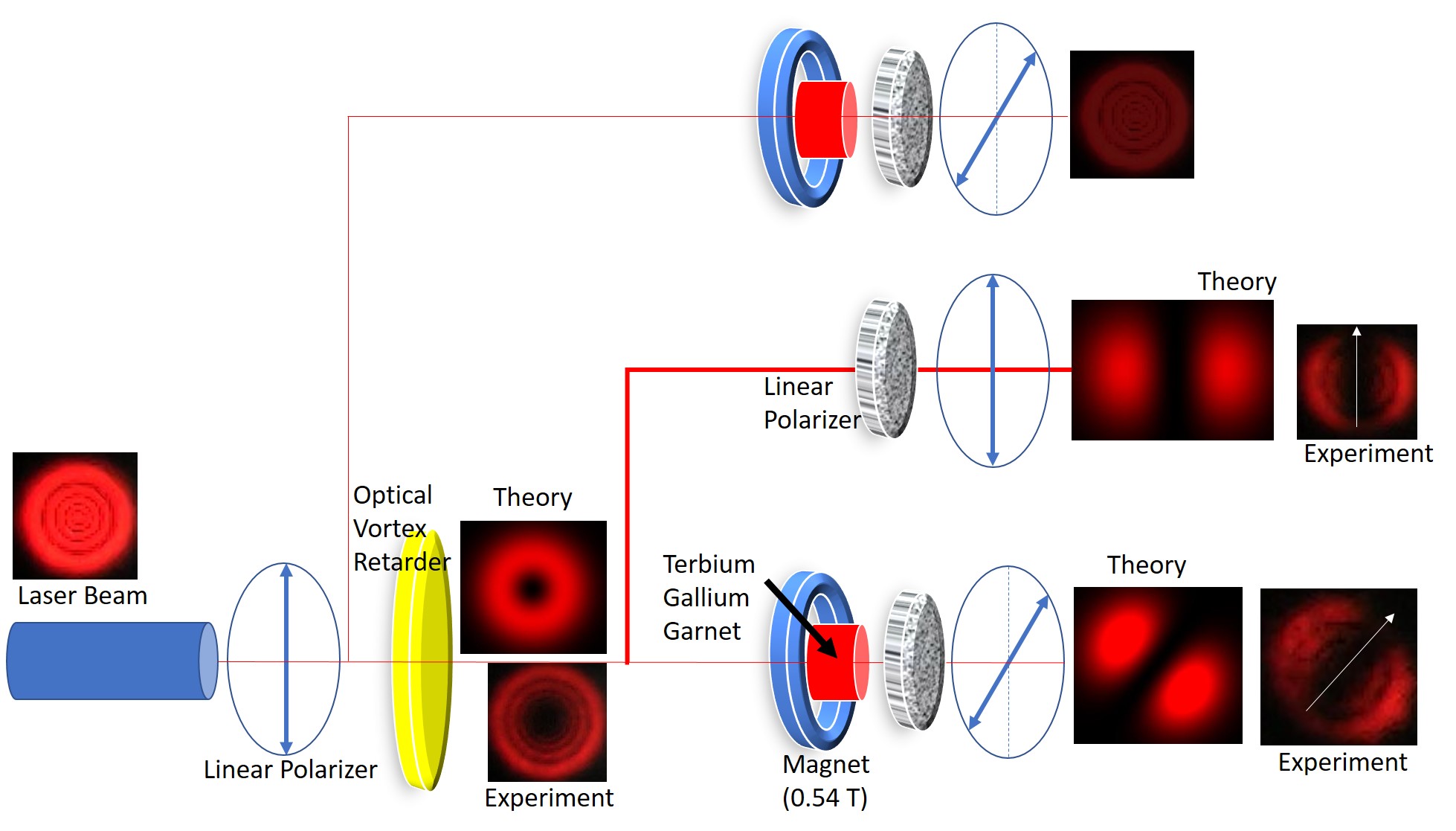}
\caption{
The experimental setup for the visible Faraday effect. Here the doughnut beam is arranged to be radially polarized. 
On the top arm shown that traditional Faraday effect setup. 
On the middle arm the doughnut intensity ring passing through a linear polarizer is transformed into two intensity lobes with the orientation of the gap between the lobes pointing vertically for this radially-polarized original LG beam. 
The bottom arm includes the magnet and the material leading to the Faraday rotation. A similar set up (not shown) with the azimuthally-polarised beam produces two lopes with the gap oriented horizontally in the middle arm and a Faraday rotation of the lobes pattern on the bottom arm.
}
\label{fig:setup}
\end{figure*}

Equation (\ref{vecxrad1}) is the analytical result for the intensity which corresponds to the experimental setup in (Fig. \ref{fig:setup}).   Its distribution on the transverse plane is shown in Fig. \ref{fig:theoretical1} in which the changes in the intensity pattern demonstrates the Faraday rotation on varying the angle of rotation $\Theta_F$. 

One side of the experimental set up is represented by the upper arm of the diagram in Fig. \ref{fig:setup}.  Here the radially-polarised LG beam passes through the second polarizer. The result of this is that the doughnut shape of the LG beam is split into two lobes separated by a dark intensity gap.  The orientation of the intensity gap is perpendicular to the polarization direction for the radially-polarized case.


The second side of the Faraday rotation set up is shown on the lower arm of Fig. \ref{fig:setup}. The figure shows the LG beam emerging from the optical vortex retarder and this beam is sent through the terbium gallium garnet sample which is subject to an axial magnetic field. The role of the material sample is to rotate the polarization direction by $45$ degrees. The linear polarizer after the sample is used to produce the splitting and observe the change in polarization with the naked eye.

In our experiment as described in Fig. \ref{fig:setup}, our aim was to show that the Faraday rotation can be directly observed by the naked eye. The essential features of the experimental set up are as follows. The laser source used has a wavelength of $640$ nm. The laser beam was first passed through the linear polarizer. It is then passed through the optical vortex lens WPV$10$L-$633$ which converts the linearly polarized laser beam into the radially or azimuthally polarized (doughnut mode) LG beam with topological charge $\ell=1$.  The doughnut beam profile is shown in in Fig. \ref{fig:sub-firsta}.  

The first part of the experiment concerns the upper arm in Fig. \ref{fig:setup}. The radially-polarised beam then passes through another linear polarizer which 
results in splitting the ring-shaped doughnut intensity distribution into two intensity lobes where the intensity gap perpendicular to the polarization direction for a radially polarized vortex retarder. The profile of the intensity of the two lobes is shown in  Fig. \ref{fig:sub-seconda} for the originally radially polarised beam. 
The set  up includes a magnetic field which is applied to the polarization sensitive material. This compact Faraday rotator includes Terbium Gallium Garnet (TGG) as the material on which a constant magnetic field of magnitude $0.54$ T is applied. Its Verdet constant is  $-131$  rad/(T·m) at the laser wavelength of $640$ $nm$ and it leads to $45$ degree rotation of the polarization direction. 
In conventional Faraday rotation experiments
  the polarization rotation is deduced from the recorded change of the intensity of the light. A major advantage of our technique, which employs  twisted light is that its results lead to the direct and easy measurement of the visible rotation.

 The same set up was used to measure change in the polarization rotation due to the constant applied magnetic field but with flint glass as the material. 
 The rotation angle was again measured as $45$ degrees, and once again observed directly as the angle of the split between two intensity lobes.


In conclusion, we have put forward and explored the workings of a new techniques for the realisation of the optical Faraday rotation experiments using radially polarized optical vortex beams.This technique exhibits the Faraday effect directly, without further steps to detect changes in the light intensity.  The degree of rotation of the plane of polarisation is readily  determined by the visibly clear change in the orientation of the split between visibly clear intensity lobes.
\subsection*{Acknowledgements}
K.K and F.T. wish to thank Bitlis Eren University for financial support (under the project:BEBAP 2021.02).
\bibliography{References}

\begin{thebibliography}{16}%
\makeatletter
\providecommand \@ifxundefined [1]{%
 \@ifx{#1\undefined}
}%
\providecommand \@ifnum [1]{%
 \ifnum #1\expandafter \@firstoftwo
 \else \expandafter \@secondoftwo
 \fi
}%
\providecommand \@ifx [1]{%
 \ifx #1\expandafter \@firstoftwo
 \else \expandafter \@secondoftwo
 \fi
}%
\providecommand \natexlab [1]{#1}%
\providecommand \enquote  [1]{``#1''}%
\providecommand \bibnamefont  [1]{#1}%
\providecommand \bibfnamefont [1]{#1}%
\providecommand \citenamefont [1]{#1}%
\providecommand \href@noop [0]{\@secondoftwo}%
\providecommand \href [0]{\begingroup \@sanitize@url \@href}%
\providecommand \@href[1]{\@@startlink{#1}\@@href}%
\providecommand \@@href[1]{\endgroup#1\@@endlink}%
\providecommand \@sanitize@url [0]{\catcode `\\12\catcode `\$12\catcode
  `\&12\catcode `\#12\catcode `\^12\catcode `\_12\catcode `\%12\relax}%
\providecommand \@@startlink[1]{}%
\providecommand \@@endlink[0]{}%
\providecommand \url  [0]{\begingroup\@sanitize@url \@url }%
\providecommand \@url [1]{\endgroup\@href {#1}{\urlprefix }}%
\providecommand \urlprefix  [0]{URL }%
\providecommand \Eprint [0]{\href }%
\providecommand \doibase [0]{http://dx.doi.org/}%
\providecommand \selectlanguage [0]{\@gobble}%
\providecommand \bibinfo  [0]{\@secondoftwo}%
\providecommand \bibfield  [0]{\@secondoftwo}%
\providecommand \translation [1]{[#1]}%
\providecommand \BibitemOpen [0]{}%
\providecommand \bibitemStop [0]{}%
\providecommand \bibitemNoStop [0]{.\EOS\space}%
\providecommand \EOS [0]{\spacefactor3000\relax}%
\providecommand \BibitemShut  [1]{\csname bibitem#1\endcsname}%
\let\auto@bib@innerbib\@empty
\bibitem [{\citenamefont {Vojna}\ \emph
  {et~al.}(2020{\natexlab{a}})\citenamefont {Vojna}, \citenamefont {Duda},
  \citenamefont {Yasuhara}, \citenamefont {Slez\'{a}k}, \citenamefont
  {Schlichting}, \citenamefont {Stevens}, \citenamefont {Chen}, \citenamefont
  {Lucianetti},\ and\ \citenamefont {Mocek}}]{vojna2020}%
  \BibitemOpen
  \bibfield  {author} {\bibinfo {author} {\bibfnamefont {D.}~\bibnamefont
  {Vojna}}, \bibinfo {author} {\bibfnamefont {M.}~\bibnamefont {Duda}},
  \bibinfo {author} {\bibfnamefont {R.}~\bibnamefont {Yasuhara}}, \bibinfo
  {author} {\bibfnamefont {O.}~\bibnamefont {Slez\'{a}k}}, \bibinfo {author}
  {\bibfnamefont {W.}~\bibnamefont {Schlichting}}, \bibinfo {author}
  {\bibfnamefont {K.}~\bibnamefont {Stevens}}, \bibinfo {author} {\bibfnamefont
  {H.}~\bibnamefont {Chen}}, \bibinfo {author} {\bibfnamefont {A.}~\bibnamefont
  {Lucianetti}}, \ and\ \bibinfo {author} {\bibfnamefont {T.}~\bibnamefont
  {Mocek}},\ }\href {\doibase 10.1364/OL.387911} {\bibfield  {journal}
  {\bibinfo  {journal} {Opt. Lett.}\ }\textbf {\bibinfo {volume} {45}},\
  \bibinfo {pages} {1683} (\bibinfo {year} {2020}{\natexlab{a}})}\BibitemShut
  {NoStop}%
\bibitem [{\citenamefont {Vleugels}\ \emph {et~al.}(2019)\citenamefont
  {Vleugels}, \citenamefont {Steverlynck}, \citenamefont {Brullot},
  \citenamefont {Koeckelberghs},\ and\ \citenamefont {Verbiest}}]{vle2019}%
  \BibitemOpen
  \bibfield  {author} {\bibinfo {author} {\bibfnamefont {R.}~\bibnamefont
  {Vleugels}}, \bibinfo {author} {\bibfnamefont {J.}~\bibnamefont
  {Steverlynck}}, \bibinfo {author} {\bibfnamefont {W.}~\bibnamefont
  {Brullot}}, \bibinfo {author} {\bibfnamefont {G.}~\bibnamefont
  {Koeckelberghs}}, \ and\ \bibinfo {author} {\bibfnamefont {T.}~\bibnamefont
  {Verbiest}},\ }\href {\doibase 10.1021/acs.jpcc.9b00607} {\bibfield
  {journal} {\bibinfo  {journal} {The Journal of Physical Chemistry C}\
  }\textbf {\bibinfo {volume} {123}},\ \bibinfo {pages} {9382} (\bibinfo {year}
  {2019})},\ \Eprint
  {http://arxiv.org/abs/https://doi.org/10.1021/acs.jpcc.9b00607}
  {https://doi.org/10.1021/acs.jpcc.9b00607} \BibitemShut {NoStop}%
\bibitem [{\citenamefont {Schatz}\ and\ \citenamefont
  {McCaffery}(1969)}]{schatz1969faraday}%
  \BibitemOpen
  \bibfield  {author} {\bibinfo {author} {\bibfnamefont {P.}~\bibnamefont
  {Schatz}}\ and\ \bibinfo {author} {\bibfnamefont {A.}~\bibnamefont
  {McCaffery}},\ }\href@noop {} {\bibfield  {journal} {\bibinfo  {journal}
  {Quarterly Reviews, Chemical Society}\ }\textbf {\bibinfo {volume} {23}},\
  \bibinfo {pages} {552} (\bibinfo {year} {1969})}\BibitemShut {NoStop}%
\bibitem [{\citenamefont {Bennett}\ and\ \citenamefont
  {Stern}(1965)}]{bennett1965faraday}%
  \BibitemOpen
  \bibfield  {author} {\bibinfo {author} {\bibfnamefont {H.~S.}\ \bibnamefont
  {Bennett}}\ and\ \bibinfo {author} {\bibfnamefont {E.~A.}\ \bibnamefont
  {Stern}},\ }\href@noop {} {\bibfield  {journal} {\bibinfo  {journal}
  {Physical Review}\ }\textbf {\bibinfo {volume} {137}},\ \bibinfo {pages}
  {A448} (\bibinfo {year} {1965})}\BibitemShut {NoStop}%
\bibitem [{\citenamefont {Serber}(1932)}]{serber1932theory}%
  \BibitemOpen
  \bibfield  {author} {\bibinfo {author} {\bibfnamefont {R.}~\bibnamefont
  {Serber}},\ }\href@noop {} {\bibfield  {journal} {\bibinfo  {journal}
  {Physical Review}\ }\textbf {\bibinfo {volume} {41}},\ \bibinfo {pages} {489}
  (\bibinfo {year} {1932})}\BibitemShut {NoStop}%
\bibitem [{\citenamefont {Boswarva}\ \emph {et~al.}(1962)\citenamefont
  {Boswarva}, \citenamefont {Howard},\ and\ \citenamefont
  {Lidiard}}]{boswarva1962faraday}%
  \BibitemOpen
  \bibfield  {author} {\bibinfo {author} {\bibfnamefont {I.~M.}\ \bibnamefont
  {Boswarva}}, \bibinfo {author} {\bibfnamefont {R.}~\bibnamefont {Howard}}, \
  and\ \bibinfo {author} {\bibfnamefont {A.}~\bibnamefont {Lidiard}},\
  }\href@noop {} {\bibfield  {journal} {\bibinfo  {journal} {Proceedings of the
  Royal Society of London. Series A. Mathematical and Physical Sciences}\
  }\textbf {\bibinfo {volume} {269}},\ \bibinfo {pages} {125} (\bibinfo {year}
  {1962})}\BibitemShut {NoStop}%
\bibitem [{\citenamefont {Vojna}\ \emph {et~al.}(2019)\citenamefont {Vojna},
  \citenamefont {Slez{\'a}k}, \citenamefont {Lucianetti},\ and\ \citenamefont
  {Mocek}}]{vojna2019verdet}%
  \BibitemOpen
  \bibfield  {author} {\bibinfo {author} {\bibfnamefont {D.}~\bibnamefont
  {Vojna}}, \bibinfo {author} {\bibfnamefont {O.}~\bibnamefont {Slez{\'a}k}},
  \bibinfo {author} {\bibfnamefont {A.}~\bibnamefont {Lucianetti}}, \ and\
  \bibinfo {author} {\bibfnamefont {T.}~\bibnamefont {Mocek}},\ }\href@noop {}
  {\bibfield  {journal} {\bibinfo  {journal} {Applied Sciences}\ }\textbf
  {\bibinfo {volume} {9}},\ \bibinfo {pages} {3160} (\bibinfo {year}
  {2019})}\BibitemShut {NoStop}%
\bibitem [{\citenamefont {Vojna}\ \emph
  {et~al.}(2020{\natexlab{b}})\citenamefont {Vojna}, \citenamefont
  {Slez{\'a}k}, \citenamefont {Yasuhara}, \citenamefont {Furuse}, \citenamefont
  {Lucianetti},\ and\ \citenamefont {Mocek}}]{vojna2020faraday}%
  \BibitemOpen
  \bibfield  {author} {\bibinfo {author} {\bibfnamefont {D.}~\bibnamefont
  {Vojna}}, \bibinfo {author} {\bibfnamefont {O.}~\bibnamefont {Slez{\'a}k}},
  \bibinfo {author} {\bibfnamefont {R.}~\bibnamefont {Yasuhara}}, \bibinfo
  {author} {\bibfnamefont {H.}~\bibnamefont {Furuse}}, \bibinfo {author}
  {\bibfnamefont {A.}~\bibnamefont {Lucianetti}}, \ and\ \bibinfo {author}
  {\bibfnamefont {T.}~\bibnamefont {Mocek}},\ }\href@noop {} {\bibfield
  {journal} {\bibinfo  {journal} {Materials}\ }\textbf {\bibinfo {volume}
  {13}},\ \bibinfo {pages} {5324} (\bibinfo {year}
  {2020}{\natexlab{b}})}\BibitemShut {NoStop}%
\bibitem [{\citenamefont {Jenkins}\ and\ \citenamefont
  {White}(1957)}]{jenkins1957fundamentals}%
  \BibitemOpen
  \bibfield  {author} {\bibinfo {author} {\bibfnamefont {F.~A.}\ \bibnamefont
  {Jenkins}}\ and\ \bibinfo {author} {\bibfnamefont {H.~E.}\ \bibnamefont
  {White}},\ }\href@noop {} {\bibfield  {journal} {\bibinfo  {journal} {Indian
  Journal of Physics}\ }\textbf {\bibinfo {volume} {25}},\ \bibinfo {pages}
  {265} (\bibinfo {year} {1957})}\BibitemShut {NoStop}%
\bibitem [{\citenamefont {Allen}\ \emph {et~al.}(1999)\citenamefont {Allen},
  \citenamefont {Padgett},\ and\ \citenamefont {Babiker}}]{allen1999}%
  \BibitemOpen
  \bibfield  {author} {\bibinfo {author} {\bibfnamefont {L.}~\bibnamefont
  {Allen}}, \bibinfo {author} {\bibfnamefont {M.}~\bibnamefont {Padgett}}, \
  and\ \bibinfo {author} {\bibfnamefont {M.}~\bibnamefont {Babiker}},\ }in\
  \href {\doibase 10.1016/S0079-6638(08)70391-3} {\emph {\bibinfo {booktitle}
  {Progress in Optics}}},\ Vol.~\bibinfo {volume} {39}\ (\bibinfo  {publisher}
  {Insitutte of Physics},\ \bibinfo {year} {1999})\ pp.\ \bibinfo {pages}
  {291--372}\BibitemShut {NoStop}%
\bibitem [{\citenamefont {Andrews}\ and\ \citenamefont
  {Babiker}(2012)}]{andrews2012}%
  \BibitemOpen
  \bibinfo {editor} {\bibfnamefont {D.~L.}\ \bibnamefont {Andrews}}\ and\
  \bibinfo {editor} {\bibfnamefont {M.}~\bibnamefont {Babiker}},\ eds.,\
  \href@noop {} {\emph {\bibinfo {title} {{The Angular Momentum of Light}}}}\
  (\bibinfo  {publisher} {Cambridge University Press},\ \bibinfo {address}
  {Cambridge},\ \bibinfo {year} {2012})\BibitemShut {NoStop}%
\bibitem [{\citenamefont {Loeffler}(1983)}]{loeffler1983faraday}%
  \BibitemOpen
  \bibfield  {author} {\bibinfo {author} {\bibfnamefont {F.~J.}\ \bibnamefont
  {Loeffler}},\ }\href@noop {} {\bibfield  {journal} {\bibinfo  {journal}
  {American Journal of Physics}\ }\textbf {\bibinfo {volume} {51}},\ \bibinfo
  {pages} {661} (\bibinfo {year} {1983})}\BibitemShut {NoStop}%
\bibitem [{\citenamefont {Pedrotti}\ and\ \citenamefont
  {Bandettini}(1990)}]{pedrotti1990faraday}%
  \BibitemOpen
  \bibfield  {author} {\bibinfo {author} {\bibfnamefont {F.~L.}\ \bibnamefont
  {Pedrotti}}\ and\ \bibinfo {author} {\bibfnamefont {P.}~\bibnamefont
  {Bandettini}},\ }\href@noop {} {\bibfield  {journal} {\bibinfo  {journal}
  {American Journal of Physics}\ }\textbf {\bibinfo {volume} {58}},\ \bibinfo
  {pages} {542} (\bibinfo {year} {1990})}\BibitemShut {NoStop}%
\bibitem [{\citenamefont {Berman}(2010)}]{berman2010optical}%
  \BibitemOpen
  \bibfield  {author} {\bibinfo {author} {\bibfnamefont {P.}~\bibnamefont
  {Berman}},\ }\href@noop {} {\bibfield  {journal} {\bibinfo  {journal}
  {American Journal of Physics}\ }\textbf {\bibinfo {volume} {78}},\ \bibinfo
  {pages} {270} (\bibinfo {year} {2010})}\BibitemShut {NoStop}%
\bibitem [{\citenamefont {Olvera-Santamar{\'\i}a}\ \emph
  {et~al.}(2020)\citenamefont {Olvera-Santamar{\'\i}a}, \citenamefont
  {Garc{\'\i}a-Garc{\'\i}a}, \citenamefont {Tlapale-Aguilar}, \citenamefont
  {Silva-Barranco}, \citenamefont {Rickenstorff-Parrao},\ and\ \citenamefont
  {Ostrovsky}}]{olvera2020cylindrically}%
  \BibitemOpen
  \bibfield  {author} {\bibinfo {author} {\bibfnamefont {M.}~\bibnamefont
  {Olvera-Santamar{\'\i}a}}, \bibinfo {author} {\bibfnamefont {J.}~\bibnamefont
  {Garc{\'\i}a-Garc{\'\i}a}}, \bibinfo {author} {\bibfnamefont
  {A.}~\bibnamefont {Tlapale-Aguilar}}, \bibinfo {author} {\bibfnamefont
  {J.}~\bibnamefont {Silva-Barranco}}, \bibinfo {author} {\bibfnamefont
  {C.}~\bibnamefont {Rickenstorff-Parrao}}, \ and\ \bibinfo {author}
  {\bibfnamefont {A.}~\bibnamefont {Ostrovsky}},\ }\href@noop {} {\bibfield
  {journal} {\bibinfo  {journal} {Optics Communications}\ }\textbf {\bibinfo
  {volume} {467}},\ \bibinfo {pages} {125693} (\bibinfo {year}
  {2020})}\BibitemShut {NoStop}%
\bibitem [{Note1()}]{Note1}%
  \BibitemOpen
  \bibinfo {note} {Thorlabs zero-order vortex half-wave retarder, model
  WPV$10$}\BibitemShut {NoStop}%
\end{thebibliography}%

\end{document}